\documentclass[12pt]{article}
\usepackage{graphicx}
\begin{document}

\centerline{\bf {\Large Evolutionary ecology in-silico: Does mathematical modelling}} 
\centerline{\bf {\large help in understanding the ``generic'' trends?}}

\bigskip 

\bigskip 

\centerline{\bf {\large Debashish Chowdhury$^{1,\ast}$ and Dietrich Stauffer$^{2,\dag}$}}

\bigskip 

\bigskip 

\centerline{$^1$ Physics Department, Indian Institute of Technology, Kanpur 208016, India.} 

\centerline{$^2$ Institute for Theoretical Physics, University of Cologne, 50923 K\"oln, Germany.} 

\bigskip 

\bigskip

\noindent{\bf Abstract: } 

Motivated by the results of recent laboratory experiments (Yoshida et al. 
Nature, 424, 303-306 (2003)) as well as many earlier field observations 
that evolutionary changes can take place in ecosystems over relatively 
short ecological time scales, several ``unified'' mathematical models of 
evolutionary ecology have been developed over the last few years with 
the aim of describing the statistical properties of data related to the 
evolution of ecosystems. Moreover, because of the availability of 
sufficiently fast computers, it has become possible to carry out detailed 
computer simulations of these models. For the sake of completeness and 
to put these recent developments in the proper perspective, we begin with 
a brief summary of some older models of ecological phenomena and 
evolutionary processes. However, the main aim of this article is to review 
critically these ``unified'' models, particularly those published in the 
physics literature, in simple language that makes the new theories 
accessible to wider audience. 

\vspace{6cm}

\noindent $\ast$ E-mail: debch@iitk.ac.in 

\noindent $\dag$ E-mail: stauffer@thp.uni-koeln.de

\newpage

\section{Introduction}

Enormous progress has been made in the twentieth century in the 
domain of sub-cellular and cell biology, particularly in area 
of molecular genetics and genomics. One of the  challenges of 
the twenty-first century will be to link the insight gained from 
the molecular level research on uni-cellular as well as 
multi-cellular organisms to biological research at higher levels 
of organization, namely, those at the levels of colonies, 
communities and, finally, eco-systems \cite{jackson02,kafatos}. 
Admittedly, at present, we are far from that goal. 

In traditional paleobiology, analysis of the fossil data has 
always been the  most popular way of understanding the causes 
and consequences of extinction of species as well as those of 
biotic recoveries from mass extinctions 
\cite{ehrlich81,raup86,raup91, miller98,jablonski04,erwin01}.  
Unfortunately, the available record of the history of life, 
written on stone in the form of fossils, is incomplete and 
ambiguous \cite{raup91}. Laboratory experiments have also 
played equally important role so far in ecology and evolutionary 
biology. However, an alternative enterprise seeks to recreate 
the evolution on a computer by simulating theoretical models; 
this is often referred to as in-silico experiments. 

Models are normally useful in understanding the real world. In 
principle, models can be verbal or symbolic, graphical or abstract, 
qualitative or quantitative. However, throughout this paper, by 
the term model we shall always mean mathematical models that not 
only indicate qualitative features of various quantities of interest 
but can also make quantitative predictions. Mathematical modelling 
often helps in getting insight into ecological phenomena and 
evolutionary processes \cite{murray,keshet}.  Most of the 
ecological and evolutionary models are too complicated to be 
solved analytically; for such models computer simulation is one of 
the most powerful tools of analysis.

It has been realized in recent years that ecosystems are examples of 
complex adaptive systems \cite{levin98,hartvigsen98,milne98, wu02}. 
Over the last ten years statistical physicists have used the conceptual 
toolbox of their profession for understanding some aspects of the 
dynamical evolution of eco-systems which include, for example, the 
``generic'' trends in the statistics of the data on speciation 
and extinction \cite{drosselrev,newmanrev,soletree}. Significant 
progress has been made over the last three years in developing 
detailed models that incorporate not only ecological phemomena on 
short periods of time but also evolutionary processes on longer time 
scales. In this article we present a critical overview of the current 
status of the {\it ``unified generic theories''} of evolutionary 
ecology.

Models intended to describe the spatio-temporal patterns in ecological 
and evolutionary processes must set the spatio-temporal {\it scale} 
unambiguously \cite{levin00,allen02}. For example, let us consider 
a single aerial photograph of a landscape. While the boundaries of 
the photograph determine the {\it spatial extent} of the observation, 
the size of the pixels (grain size) in the photograph imposes the 
limit  on the {\it spatial resolution}. Similarly, if a 
sequence of photographs of the same landscape is taken at regular 
intervals, the time difference between the first and the last 
photograph is a measure of the {\it temporal extent} of the 
observations while the time difference between the successive 
photographs determines the corresponding {\it temporal resolution} 
\cite{martinez1}. For example, in ecology the temporal resolution 
can be days while the temporal extent can be up to decades, whereas 
the temporal resolution in evolutionary biology, particularly 
empirical observations from fossil data varies, usually, from tens 
of thousands to millions of years. In this review we shall consider 
different classes of models with widely different scales of 
spatio-temporal resolution.

The ``ecological'' models, that describe population dynamics in detail
using, for example, the Lotka-Volterra equations (discussed in section 
\ref{sec3}) usually ignore the slow macro-evolutionary changes in the 
eco-system; hardly any effects of these would be observable before the 
computer simulations would run out of computer time. On the other hand, 
in order to simulate the billion-year old history of life on earth with 
a computer, the elementary time steps in ``evolutionary'' models have 
to correspond to thousands of years, if not millions; consequently, the 
finer details of the ecological processes over shorter periods of time  
cannot be accounted for by these models in any explicit manner. However, 
despite the practical difficulties, it is desirable, at least in 
principle, to develop one single theoretical model which would be 
able to describe the entire dynamics of an eco-system since the first 
appearance of life in it up till now and in as much detail as possible. 
This dream has now come closer to reality, mainly because of the 
availability of fast computers. It has now become feasible to carry out 
computer simulations of eco-system models where, each time step (i.e., 
temporal {\it resolution}) would correspond to typical times for 
``micro''- evolution while the total duration (i.e., temporal 
{\it duration}) of each of the simulations is long enough to capture 
``macro''-evolution.

\begin{figure}[h]
\begin{center}
\includegraphics[scale=0.50,angle=-90]{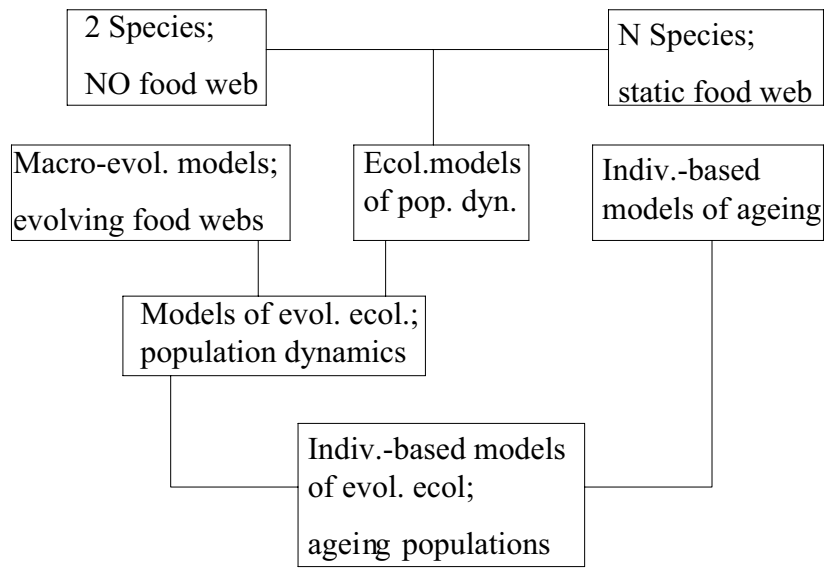}
\end{center}
\caption{Broad classification of the mathematical models of 
evolutionary ecology with different levels of detailed description. 
}
\label{fig-table}
\end{figure}

The mathematical models in evolutionary ecology can be broadly 
classified into different classes with different levels of detailed 
description, as shown in Fig.\ref{fig-table}. 


In the earliest mathematical models of population dynamics, only 
one predator species and one prey species were considered. However,  
for modelling the population dynamics of more than two species, one
needs to know the {\it food web} which is a graphical way of
describing the prey-predator relations, i.e., which species eats
which one and which compete among theselves for the same food
resources \cite{pimm,polis,drossel03,cohen90}. More precisely, a
food web is a directed graph where each node is labelled by a
species' name and each directed link indicates the direction of
flow of nutrient (i.e., {\it from} a prey {\it to} one of its
predators). In the early works of this type, the food web was 
assumed to be static, i.e., independent of time. An altogether 
different class of models were developed to study macro-evolution; 
in such models, because of the Darwinian evolution, the food web 
is a dynamic network. In recent times, models of evolutionary 
ecology have been developed by a synthesis of ecological models 
of population dynamics and macroevolutionary models with evolving 
food webs. However, a more detailed theoretical description has 
also been attempted by incorporating individual organisms explicitly 
in the model where the birth, ageing and death of each individual 
occurs naturally.


\vspace{1cm} 

Almost all the models developed along the lines of models of physical 
systems usually reach a stationary state after sufficiently long time. 
In contrast to these models, the models closer to biological reality 
might never reach a stationary state. After all, life forms in nature 
have evolved over billions of years from simple bacteria and archaea 
to complex structures like the bodies of dinosaurs and the brains of 
our readers. 

So far as the results of theoretical modeling are concerned, most 
of the recent works in physics literature have been concerned with 
the possibility of the existence of ``power laws'' in the statistics 
of extinction data. For example, suppose it is claimed that the 
relative frequency $P(s)$ of extinctions of size $s$ follows the 
power-law $P(s) \propto s^{-\tau}$ with $\tau \simeq 2$. If irrefutable 
evidences in favour of such power laws can be gathered, either from 
fossil data or from mathematical modeling, it would imply that the 
self-organizing dynamics of eco-systems exhibit fluctuations that are 
statistically self-similar because a change of scale ($s \to s' = b s$) 
leaves the form of the power law unchanged. 

\section{Ageing and age-structured population in single species} 

Biological ageing of adults is best measured through the mortality
$q(a) = [S(a)-S(a+1)]/S(a)$ where $S(a)$ is the number of survivors 
to the age $a$ in suitable time units (like years for humans or 
days for flies and worms). More accurate is the
mortality function $\mu(a) = - d \ln S(a)/da$ which is defined in terms 
of a derivative instead of a difference. The Gompertz law states that the
mortality function for adults increases exponentially with age; this is also
valid for many animals, but does not hold for the youngest and perhaps the
oldest ages \cite{vaupel}. Many theories exist to explain ageing 
\cite{finch90,rose94,muellerrose}
:accumulation of hereditary detrimental mutations \cite{moss}, reliability 
theory as known from engineering in inanimate machines \cite{gavrilov}, 
loss of telomeres in cell division \cite{aviv,joeng}, damage caused by 
oxygen radicals, trade-off between longevity and fecundity (disposable 
soma), wear and tear, etc. Only
for the first three, several quantitative computer simulations or mathematical
solutions, giving the Gompertz law, are known to us, as listed in the cited
literature.

These models do not explicitly incorporate inter-species interactions 
like, for example, prey-predator interactions. These models cannot 
capture macro-evolutionary phenomena like, for example, extinctions 
which depend crucially on the prey-predator interactions. Such 
interactions are an essential part of the ecological models which we 
mention briefly in the next section.

\section{\label{sec3} Ecological models of population dynamics: prey-predator interactions} 

Traditionally, the population dynamics of prey-predator systems 
have been described quantitatively in terms of the Lotka-Volterra 
equations \cite{goel71a,goel71b,pielou77,emlen84}. The nonlinearity 
of these deterministic differential equations leads to a rich 
variety of dynamical behaviour of the system. Although originally 
only two interacting species (predator and prey) were considered, 
later the approach was extended to more than two (but only a few) 
interacting species. Pioneering mathematical work of May 
\cite{may74} raised the question of stability of these dynamical 
equations when the number of interacting species increases. This 
challenged the earlier common belief that diversity of species 
ensured enhanced stability of the eco-system 
\cite{svirezhev83,logofet93,hofbauer98,mccann00,sinha2}.

Unlike the models of ageing discussed in the preceeding section, 
these Lotka-Volterra-type models of population dynamics monitor only 
the increase (or decrease) of populations caused by the birth (or 
death) of individual organisms but, usually, do not keep track of 
their ageing with time. 

The original formulation of the Lotka-Volterra equations assume 
that the population of the prey as well as that of predators are 
uniformly distributed in space. The absence of the spatial degrees 
of freedom in these equations is usually interpreted in the 
statistical physics literature as a mean-field-like approximation. 
This situation is similar to a well-stirred chemical reaction 
where the spatial fluctuations in the concentrations of the 
reactants and the products is negligibly small. On the other 
hand, spatial inhomogeneities in the eco-systems and migration 
of organisms from one eco-system to another are known to play 
crucial roles in evolutionary ecology \cite{czaran,bascompte,tilman}.

In recent years, the spatial inhomogeneity of the populations, i.e., 
variation in the population of the same spacies from one spatial 
patch to another \cite{singh}, that have been observed in real ecosystems has  
been captured by extending the Lotka-Volterra systems on discrete 
lattices where each of the lattice sites represents different 
spatial patches or habitats of the ecosystem
\cite{tainaka89,satulovsky94,boccara94,frachebourg96,lipowski99,
antal01a,antal01b,droz02,droz04,johst99}.

For modelling the population dynamics of more than two species, one 
needs to know the {\it food web} which is a graphical way of 
describing the prey-predator relations, i.e., which species eats 
which one and which compete among theselves for the same food 
resources \cite{pimm,polis,drossel03,cohen90}. 
The structure of foodwebs and their statistical 
properties have been investigated both using field data on real 
eco-systems as well as abstract mathematical models 
\cite{martinez1,hall91,goldwasser93,martinez95a,martinez00,
martinez02,martinez02a,martinez04,briand84,cohenbook,montoya01,
montoya02,jennings03}, particularly, in the recent years in the light 
of scale-free and small world networks \cite{strogatz01,barabasi02}. 

A static (time-independent) food web may be a good 
approximation over a short period of time. But, a more 
realistic description, valid over longer period of time, must 
take into account not only the adaptations of the species and 
their changing food habits, but also their extinction and 
creation of new species through speciation or migration of 
alien species into a new habitat. These processes make 
the food web a slowly evolving graph. Such slow time evolution 
of the food webs are naturally incorporated in macroevolutionary 
models which we summarize in the next section.

\section{Modelling macroevolution and extinction: evolving food webs}

These models are intended to throw light on the mechanisms of 
origination of species through speciation as well as their 
extinction arising from biotic and abiotic causes. Several models 
have been developed just to account for the different routes to 
speciation \cite{schluter00,diekmann99,doebeli00,kirkpatrick02,
gavrilets98,gavrilets99,gavrilets00,gavrilets03,kaneko00}. However, in this 
section we shall focus mainly on those works that have been 
inspired by close similarity with concepts or phenomena in 
statistical physics.

\subsection{Self-organized critical models of eco-systems}

Inspired by the work of Per Bak and Kim Sneppen \cite{baksneppen}, 
a large number of evolutionary models have been developed over 
the last ten years 
\cite{baksneppen,paczuski96,newsnep96a,newsnep96b,kramer96,vande96,head97,solerev,soleman1,soleman2,manpac96,solebas96,soleetal96,soleetal97,roberts96,newman96,newman97,newman00,wilke97} 
Most of these works, including that of Bak and Sneppen, claimed 
the discovery of self-organized criticality \cite{soc} in the 
statistics of the numerical data on extinction. They also drew 
attention to the close relation of these observations with the 
concept of ``punctuated equilibrium'' introduced many years ago, 
by Gould and Eldredge \cite{gould1,gould2} in 
the context of extinction of species \cite{boettcher}.

\subsection{Modelling evolution as a walk on a fitness landscape}

An alternative approach views macroevolution as random walk in a 
rugged ``fitness landscape'' \cite{kauffman}. In recent years 
this approach has been extended by allowing slow evolution of 
the landscape itself to incorporate the effect of co-evolution of 
species \cite{peliti1,peliti2,wilkerev}. The notion of ``fitness'' 
has been used often loosely to mean different things \cite{brookfield}.

\subsection{Modelling eco-system as network of interacting species}

A network model of ecosystems was developed by Sole and Manrubia 
\cite{soleman1,soleman2}.
The system consists of $N$ species, each labelled by an index $i$
($i = 1,2,...N$). The state of the $i$-th species is represented
by a two-state variable $S_i$; $S_i = 0$ or $1$ depending on whether
it is extinct or alive, respectively. The inter-species interactions
are captured by the interaction matrix ${\bf J}$; the element
$J_{ij}$ denotes the influence {\it of} the species $j$ {\it on}
the species $i$.  If $J_{ij} > 0$ while, simultaneously, $J_{ji} < 0$
then $i$ is the predator and $j$ is the prey. On the other hand, if
both $J_{ij}$ and $J_{ji}$ are positive (negative) the two
species cooperate (compete).
                                                                                
The dynamics of the system consists in updating the states of the
system (i.e., to determine the state at the time step $t+1$ from a
complete knowledge of the state at the time $t$) in the following
three steps:\\
                                                                                
\noindent {\it Step (i)}: One of the input connections $J_{ij}$ for
each species $i$ is picked up randomly and assigned a new value
drawn from the uniform distribution in the interval $[-1,1]$,
irrespective of its previous magnitude and sign (this, we believe, 
is not a very realistic description of the inter-species interaction).\\
                                                                                
\noindent {\it Step (ii)}: The new state of each of the species is
decided by the equation
\begin{equation}
S_i(t+1) = \Theta \biggl(\sum_{j=1}^N J_{ij} S_j(t) - \theta_i\biggr)
\end{equation}
where $\theta_i$ is a threshold parameter for the species $i$ and
$\Theta(x)$ is the standard step function, i.e., $\Theta(x) = 1$
if $x > 0$ but zero otherwise. If $S(t+1)$ becomes zero for $m$
species, then an extinction of size $m$ is said to have taken
place.\\
                                                                                
\noindent {\it Step (iii)}: All the niches left vacant by the extinct
species are refilled by copies of one of the randomly selected
non-extinct species.\\
                                                                                
Sole and Manrubia \cite{soleman1,soleman2} recorded extinctions of 
sizes as large as $500$ and the distributions of the sizes of these 
extinctions could be fitted to a power law of the form 
$N(m) \sim m^{-\alpha}$ with an exponent $\alpha \simeq 2.3$. Moreover, 
the periods of stasis $t_s$ were also found to obey a power law 
$N(t_s) \sim t_s^{-\gamma}$ with the exponent $\gamma \simeq 3.0$. 
However, surprisingly, in none of their papers \cite{soleman1,soleman2}, 
did Sole and his collaborators report the distributions of the lifetimes 
of species which, according to some claims (see, for example, refs.
\cite{drosselrev,newmanrev} for references to the experimental literature 
and data analysis), also follows a power law.

Amaral and Meyer \cite{amaral} considered a hierarchical food web 
which was assumed to be organized into trophic levels; a species 
in level $\ell$ feeds on some species at the level $\ell-1$ (except 
for those at $\ell = 1$ which are autotrophic). Origination of 
species through speciation was assumed to take place as follows: 
an empty niche is occupied by a non-extinct species at the same 
trophic level and the prey of the new species are selected randomly 
from among those at the level immediately below the trophic level 
of the new species. However, Amaral and Meyer did not treat the 
population dynamics of the species explicitly. Instead, a fraction 
$p$ of the species at the lowest level is randomly selected and 
made extinct. Then, any species in the next higher level for 
which all prey species became extinct are also made extinct; this 
procedure is repeated for all the levels upto the highest one. 
Although this may be a more realistic description of inter-species 
interactions than that in the Sole-Manrubia model, the dynamics of 
the model is oversimplified. From the point of view of mathematical 
analysis, the advantage of this model is that its properties can be 
obtained not only numerically \cite{amaral,camacho} but also 
analytically \cite{drossel98}.

The main limitation of these models i(see also \cite{sinha}) is that 
the individual organisms do not appear explicitly. On the other hand,  
it is the individual organisms, rather than species, which are the 
primary objects of selection \cite{lloyd93,mayr97,gould99,johnson02}. 
Moreover, the extinction of a species is nothing but eventual demise 
of all the individual organisms. 
Furthermore, direct experimental evidences   
\cite{thompson98a,thompson98b,thompson99,stockwell03,turchin03,
yoshida03,fussmann03} have established that significant evolutionary 
changes can occur over ecologically relevant time scales. In other 
words, the dynamics of Ecology and Evolution are inseparable.

\section{"Unified" models of evolutionary ecology}

The need for ``unification'' of the various ecological subdisciplines, 
e.g., population ecology, community ecology and evolutionary ecology, 
has been felt for quite some time \cite{martinez95}. In the recent 
years attempts have been made to model evolutionary ecology in terms 
of ``unified'' models that describe both micro- and macro- evolution. 
Some of these models describe population dynamics in terms of one 
single dynamical variable and, therefore, fail to account for the 
age-structured populations of each species. However, only in 
the last two years it has been possible to develop detailed models 
that describe the birth, ageing and death of individual organisms. 
This is partly because of the availability of relatively fast 
computers. 

Abramson \cite{abramson} considered a simple evolving ecosystem where 
each site of a one-dimensional lattice of finite length $L$ represent 
a species such that the species $i$ feeds on the species $i-1$ and 
its eaten by the species $i+1$. The species $1$, which feeds at a 
constant rate on the environment, represents the species at the lowest 
level of the hierarchy whereas the species $L$, occupying the top of 
the chain, is not eaten by any other species. Such a linear food web 
is not realistic, but the importance of the model lies in its its 
simplicity. 

A species is considered extinct when its population falls below a 
preassigned threshold. Because of the one-dimensional nature of the food 
web, the system would break into disjoint parts if any site is allowed to 
remain vacant following extinction of the corresponding species. 
In order to avoid such a situation, each niche that is left vacant by 
the extinction of a species is re-filled by another new species which 
interacts with the two neighbouring species with interactions whose 
strengths are drawn from a uniformly distributed random fraction. 
However, Abramson did not monitor the ageing of each individual organisms. 
Instead, he monitored only the time evolution of the total populations 
of each species in the eco-system.

McKane, Higgs and collaborators \cite{mckane98,mckane01,quince1,quince2} 
also modelled the population dynamics in terms of one single 
dynamical variable. But, unlike, Abramson \cite{abramson}, they 
took into account the hierarchical organization of the species 
in food webs. In their webworld model, each species is represented 
by a set of $L$ features (or, phenotypic characters) chosen from a 
set of $K$ possible features. Evolution of the webworld model of 
eco-system is implemented by speciation events during which a new 
species is created from a randomly chosen existing species; the 
new species differs from the parent species by just one randomly 
chosen feature. There are some close similarities between this 
model and some models developed in recent years incorporating the 
individual organisms explicitly in the model.

Over the last two years, a few ``unified'' models of evolutionary 
ecology have been developed incorporating the individual organisms 
explicitly \cite{chowstauprl,chowstaupre, stauchowpa,chowstaupa,csnetofnet,stauetal04, rikvold03,hall1,collobiano1}. 

Each individual organism in the Rikvold-Zia model \cite{rikvold03} 
has a {\it genome} of $L$ {\it genes}, each of which can take one of 
two possible values, namely, $0$ and $1$. Thus the total number 
of different {\it genotypes} is $2^L$. Rikvold and Zia \cite{rikvold03} 
assumed that each of the different genotypes represent a separate 
species. A plausible justification, suggested by Rikvold and Zia, 
is that each binary ``gene'' actually represents a group of real 
genes in a coarse-grained sense. The spirit in which such 
``coarse-grained genes'' are used in this model is somewhat similar 
to that of using the ``phenotypic'' characters in the webworld model 
\cite{mckane98,mckane01,quince1,quince2}. The number of individuals of 
genotype $I$ in generation $t$ is $n_I(t)$; the total population is 
$N_{tot}(t) = \sum_I n_I(t)$. 

In each generation, the genomes of the individual organisms are 
subjected to random mutation with probability $\mu/L$ per gene 
per individual where $L$ is the total size of the genome. Thus, 
by working with the genomes, Rikvold and Zia account for the 
genetic mutations explicitly and use the genotypes to label the 
different species.

In order to keep the model as simple as possible, Rikvold and 
Zia assumed the successive generations to be {\it nonoverlapping}. 
More precisely, the organisms are incapable of living through 
successive reproduction cycles; an individual organism produces 
a litter of $F$ offspring and immediately thereafter it dies.
Consequently, this model does not describe age-structured populations 
of any species.

Rikvold and Zia \cite{rikvold03} considered a random food web where 
the effects of species $j$ on the population of the species $i$ 
is modelled by the element $J_{ij}$ of the interaction matrix 
${\bf J}$, exactly as in the Sole-Manrubia model \cite{soleman1,soleman2}. 
In this model $J$ is taken to be a time-independent random matrix, 
with vanishing diagonal elements, whose off-diagonal elements are 
selected randomly from a uniform distribution over the interval 
$[-1,1]$. 

In each generation, the probability that an individual of 
genotype $I$ produces a litter of $F$ offspring before it dies 
is $P_I(\{n_J(t)\})$ whereas the probability that it dies without 
giving birth to offspring is $1 - P_I$. The reproduction probability 
$P_I$ is assumed to be given by \cite{rikvold03} 
\begin{equation}
P_I(\{n_J(t)\}) = \biggl\{1 + \exp\biggl[-\frac{\sum_J J_{IJ} n_J}{N_{tot}(t)} + \frac{N_{tot}(t)}{N_0}\biggr]\biggr\}^{-1} 
\end{equation}
Here the second term in the exponential, commonly called the 
Verhulst factor, represents an environmental carrying capacity 
where $N_0$ is determined by the limited shared resorces like, 
for example, space, water, light, etc.

The tangled nature model \cite{hall1,collobiano1} is slightly more 
general than the model studied by Rikvold and Zia because overlapping 
generations are allowed. An individual organism is removed from 
the system with a constant probability $p_{kill}$ per time step. 
The algorithm used for reproduction probability in the tangled 
nature model is also slightly different from that used by Rikvold 
and Zia. However, to our knowledge, there is no apriori justification 
at present for preferring either of these.    

In our works \cite{chowstauprl,chowstaupre,stauchowpa,chowstaupa,stauetal04}, 
we have provided the most detailed description of the ecological as 
well as the evolutionary processes. We have incorporated not only 
the hierarchical architecture of the natural food webs in a 
simplified manner but also the emergence of this architecture through 
self-organization as well as the possibility of migration of populations 
from one ``patch'' to another of the same eco-system for predation 
or merely for occupying a habitat.

\begin{figure}[h]
\begin{center}
\includegraphics[scale=0.50,angle=-90]{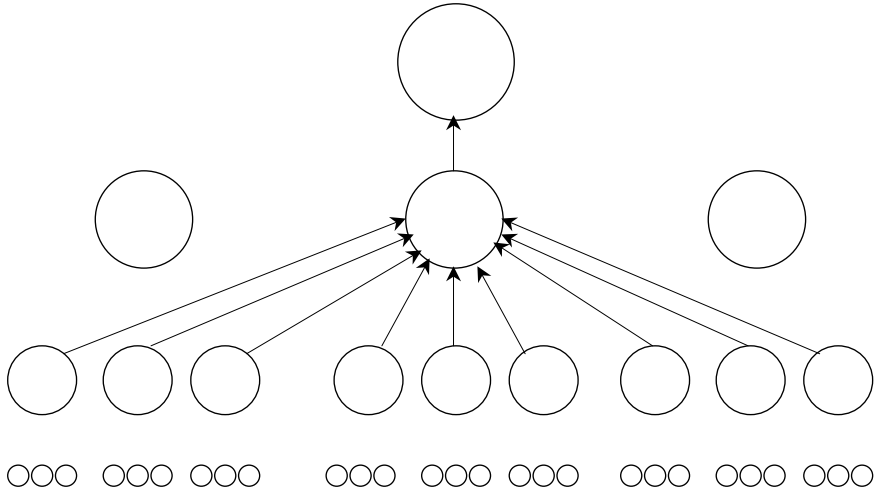}
\end{center}
\caption{A schematic representation of the network model, with {\it
hierarchical} foodweb architecture \cite{chowstaupre}. The circles 
represent the niches in the eco-system. Each arrow represents direction 
of nutrient flow.  All possible nutrient flows  {\it to} the species 
occupying the second node at the second level and that occupying the 
highest level are shown explicitly.  
}
\label{fig-hierarchy}
\end{figure}

We have modelled the eco-sytem as a dynamic {\it hierarchical} network 
(see fig.\ref{fig-hierarchy}).  Each node of the network represents 
a niche, rather than a species. Each niche can be occupied by at 
most one species at a time. The ``micro''-evolution, i.e., the birth, 
growth (ageing) and natural death of the individual organisms, in our 
model is captured by the intra-node dynamics. The ``macro``-evolution, 
e.g., adaptive co-evolution of the species, is incorporated in the 
same model through a slower evolution of the network itself over longer 
time scales. Moreover, as the model eco-system evolves with time, 
extinction of species is indicated by vanishing of the corresponding 
population; thus, the number of species and the trophic levels in the 
model eco-system can fluctuate with time. Furthermore, the natural 
process of speciation is implemented by allowing re-occupation of the 
vacant nodes by mutated versions of non-extinct species.

The prey-predator interaction between two species that occupy the
nodes $i$ and $k$ at two adjacent trophic levels is represented by
$J_{ik}$; the three possible values of $J_{ik}$ are $\pm 1$ and $0$.
The sign of $J_{ik}$ indicates the direction of trophic flow, i.e.
{\it from the lower to the higher} level. $J_{ik}$ is $+1$ if $i$
is the predator and $k$ is the prey species and it is $-1$ if $k$ 
is the predator and i denotes the prey. If there is no prey-predator
relation between the two species $i$ and $k$, we must have $J_{ik} = 0$.
This formulation of the inter-species interactions is very similar 
to that in the Sole-Manrubia model \cite{soleman1,soleman2}.
Although there is no direct interaction between species at the same
trophic level in our model, they can compete, albeit indirectly, with
each other for the same food resources available in the form of prey
at the next lower trophic level. 

The elements of the matrix $J$ account not only for the 
{\it inter}-species interactions (as in the Sole-Manrubia type models) 
but also for the {\it intra}-species interactions arising from the 
competition of individual organisms for the same food resources. In 
order to understand this interesting feature of the matrix $J$, 
consider now the two sums
\begin{equation}
{\cal S}_i^{\pm} = \pm \sum_{j=1}^N \frac{(J_{ij}^{\pm} - J_{ji})}{2} n_j
\end{equation}
where the superscript $\pm$ on $J_{ij}$ indicates that the sum is
restricted to only the positive (negative) elements $J_{ij}$. The
sum ${\cal S}_i^{+}$ is a measure of the total food {\it currently}
available to the $i$-th species whereas $-{\cal S}_i^{-}$ is a measure
of the total population of the $i$-th species that would be, at the
same time, consumed as food by its predators. If the food available
is less than the requirement, then some organisms of the species $i$
will die of {\it starvation}, even if none of them is killed by any
predator. This way the matrix ${\bf J}$ can account for the shortfall
in the food supply and the consequent competition among the organisms
of the species $i$ .

The {\it intra}-species competition among the organisms of the same 
species for limited availability of resources, other than food, 
imposes an upper limit $n_{max}$ of the allowed population of each 
species; $n_{max}$ is a time-independent parameter in the model.
Our model captures the {\it starvation deaths and killing by the
predators}, in addition to the natural death due to ageing.

\begin{figure}[h]
\begin{center}
\includegraphics[scale=0.50,angle=-90]{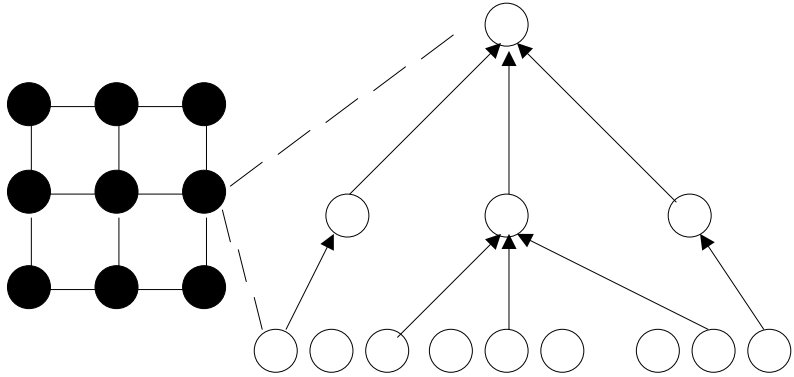}
\end{center}
\caption{The eco-system is a network (represented schematically 
by the square lattice) of spatial ``patches'' each node of which 
is endowed with a food web, another hierarchical network. 
}
\label{fig-netofnet}
\end{figure}

In our approach, each species is characterized by three features, 
namely, the minimum reproductive age $A_{rep}$, the maximum possible 
age $A_{max}$ and the litter size $M$. An organism becomes mature 
enough for reproduction only on attaining the age $A_{rep}$; beyond 
this age, the probability that it gives birth (simultaneously to 
$M$ offsprings) varies with its age $A$. The probability of its 
death due to ageing is also a function of its age $A$ but becomes 
a certainty on attaining the age $A_{max}$ provided it survives 
till then evading its predators. Although these three  
characteristics of each species are reminiscent of the features of 
the species in the webworld model \cite{mckane98,mckane01,quince1,quince2}, 
in our model these features are are not fixed parameters but are 
determined by self-organizing dynamics of the eco-system.

\begin{figure}[h]
\begin{center}
\includegraphics[scale=0.50,angle=-90]{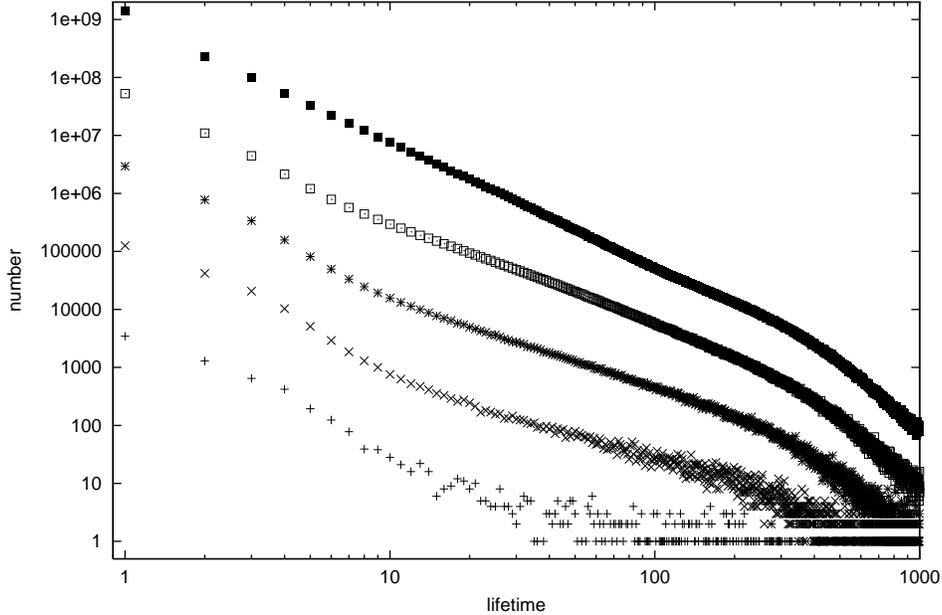}
\end{center}
\caption{Distributions of the lifetimes of the species in the latest 
version of our ``unified'' model of evolutionary ecology. The size  
of the eco-system is $5 \times 5$ in an appropriate dimensionless 
unit where each unit corresponds to a large patch. The lifetimes 
of the species are indicated along the X-axis whereas the {\it number} 
of times species with a given lifetime are encountered in our simulation, 
is plotted along the corresponding Y-axis.  The symbols 
$+$, $\times$, $\ast$, open and filled squares correspond, 
respectively, to $t = 10^3, 10^4, 10^5, 10^6$ and $10^7$, where 
$t$ is the time in dimensionless units (but, can be interpreted, for  
example, as one year) for which the ecosystem evolves following the 
dynamics of our model. 
}
\label{lifetime}
\end{figure}

In the original version of our ``unified'' model, we assumed that 
the population of the prey as well as that of predators are 
uniformly distributed in space. This situation is similar to a 
well-stirred chemical reaction where the spatial fluctuations in 
the concentrations of the reactants and the products is negligibly 
small. On the other hand, spatial inhomogeneities in the eco-systems 
and migration of organisms from one spatial ``patch'' to another 
are known to play important roles in evolutionary ecology 
\cite{czaran,bascompte,tilman}. We have captured the spatial 
inhomogeneities of the populations and characteristics of the species 
from one patch to another by extending our ``unified'' model on 
discrete lattices where each of the lattice sites represents different 
spatial ``patches'' or ``habitats'' of the eco-system \cite{stauetal04} 
Thus, the eco-system is a network of spatial ``patches'' each of 
which is endowed with a food web; in other works, the eco-system 
is a network of networks (see fig.\ref{fig-netofnet}).

As an example, in fig.\ref{lifetime} we show the distribution of the 
lifetimes of the species in the latest version of this model.
Very recently it has been pointed out, through independent computer 
simulations by Singh and Ramaswamy \cite{rama}, that the deviation 
from power-law observed in the distributions of the lifetimes in the 
``unified'' model is not any artefact of the simplified strengths 
$ J_{ij} = \pm 1$ but is a generic feature of the model with even 
more general interactions.

\section{Conclusion} 

Field studies and laboratory experiments have convincingly established 
that evolutionary changes can take place in ecosystems over relatively 
short ecological time scales 
\cite{thompson98a,thompson98b,thompson99,stockwell03,turchin03,yoshida03,fussmann03}. 
Motivated by these observations and because of the availability of 
sufficiently fast computers, several ``unified'' models of evolutionary 
ecology have been developed over the last few years 
\cite{chowstauprl,chowstaupre,stauchowpa,chowstaupa,stauetal04,rikvold03,hall1,collobiano1}. 
All of these models treat an ecosystem as a dynamically evolving 
network of species. This modelling strategy is very similar to that 
followed for modelling complex adaptive systems, an active area of 
research in statistical physics. These models provide a ``unified'' 
description of the ''generic'' features of ecology and evolution- 
ecological changes, e.g., variation of populations of different 
species, take place over relatively short time scales while 
evolutionary changes, e.g., speciation and extinction occur slowly 
over longer time scales. In this paper we have critically reviewed 
the results of computer simulations of such ``unified'' models of 
evolutionary ecology.

For the sake of completeness and to put the recent developments in 
the proper perspective, we have also mentioned briefly some earlier 
works, from simple to complicated models. Simple models have the 
advantage that they often give clear results with limited 
computational effort, like power laws for the distributions of 
lifetimes and avalanches of extinctions of species. Complicated
models are usually more realistic but, because of the lack of the 
simplicity, these may give a superposition of several different 
laws in different regimes and require enormous computational 
efforts. Nevertheless, it has been possible not only to capture 
birth, ageing of individual organisms and the prey-predator 
interactions as well as extinctions and speciation but also 
the emergence of self-organized hierarchical architecture of the 
food webs. We hope this critical review of the recent models 
published in the physics literature will stimulate interactions 
between physicists and evolutionary ecologists.

\bigskip

\noindent {\bf Acknowledgements:} We thank the anonymous referees 
for their constructive criticism as well as their valuable comments 
and suggestions. We are also indebted to the corresponding editor 
for her patience and for encouraging us to complete the project. 
This work has been supported by DFG/BMZ through a joint Indo-German 
research project.


\newpage

\begin{thebibliography}{99}

\bibitem{jackson02} R.B. Jackson, C. R. Linder, M. Lynch, M. 
Purugganan, S. Somerville and S.S. Thayer, {\it Linking molecular 
insight and ecological research}, Trends in ecology and evolution, 
{\bf 17}, 409-414 (2002).

\bibitem{kafatos} F.C. Kafatos and T. Eisner, {\it Unification in 
the century of biology}, Science {\bf 303}, 1257 (2004).

\bibitem{ehrlich81} P.R. Ehrlich and A.H. Ehrlich, {\it Extinctions: 
the causes and consequences of the disappearance of species}, 
(Ballantine, New York, 1981).

\bibitem{raup86} D.M. Raup, {\it Biological extinction in earth history}, 
Science {\bf 231}, 1528-1533 (1986). 

\bibitem{raup91} D.M. Raup, {\it Extinction-bad genes or bad luck?}, 
(WW Norton, 1991); New Scientist, {\bf 131}, 14 September issue, 36 (1991).

\bibitem{miller98} A.I. Miller, {\it Biotic transitions in global 
marine diversity}, Science {\bf 281}, 1157-1160 (1998).

\bibitem{jablonski04} D. Jablonski, {\it Extinction: past and present}, 
Nature, 427, 589 (2004).

\bibitem{erwin01} D.H. Erwin, {\it Lessons from the past: biotic 
recoveries from mass extinctions}, Proc. Natl. Acad. Sci. {\bf 98}, 
5399-5403 (2001).

\bibitem{murray} J.D. Murray {\it Mathematical Biology}, (Springer, 1989). 

\bibitem{keshet} L. Edelshtein-Keshet, {Mathematical Models In Biology}, 
(McGraw-Hill, 1988)

\bibitem{levin98} S.A. Levin, {\it Ecosystems and the biosphere as 
complex adaptive systems}, Ecosystems, {\bf 1}, 431-436 (1998). 

\bibitem{hartvigsen98} G. Hartvigsen, A. Kinzig and G. Peterson, 
{\it Use and analysis of complex adaptive systems in ecosystem 
science: overview of special section}, Ecosystems {\bf 1}, 427-430 
(1998). 

\bibitem{milne98} B.T. Milne, {\it Motivation and benefits of complex 
systems approaches in ecology}, Ecosystems, {\bf 1}, 449-456 (1998). 

\bibitem{wu02} J. Wu and D. Marceau, {\it Modeling complex ecological 
systems: an introduction}, Ecological Modelling {\bf 153}, 1-6 (2002).

\bibitem{drosselrev} B. Drossel, {\it Biological evolution and statistical 
physics}, Adv. Phys. {\bf 50}, 209-295 (2001). 
                                                                                
\bibitem{newmanrev} M. E. J. Newman, and R. G. Palmer, {\it Modeling 
Extinction}, (Oxford University Press, 2002). 

\bibitem{soletree} R.V. Sole, S.C. Manrubia, M. Benton, S. Kauffman 
and P. Bak, {\it Criticality and scaling in evolutionary ecology}, 
Trends in Ecology and Evolution, {\bf 14}, 156-160 (1999).

\bibitem{levin00} S.A. Levin, {\it Multiple scales and the maintenance 
of biodiversity}, Ecosystems {\bf 3}, 498-506 (2000).

\bibitem{allen02} C.R. Allen and C.S. Holling, {\it Cross-scale 
structure and scale breaks in ecosystems and other complex systems}, 
Ecosystems {\bf 5}, 315-318 (2002).

\bibitem{martinez1} N.D. Martinez and J.A. Dunne, {\it Time, space, 
and beyond: scale issues in food-web research}, in: {\it Ecological 
scale: theory and applications}, eds. D.L. Peterson and V. T. Parker, 
(Columbia University Press, 1998).

\bibitem{vaupel} J.W. Vaupel et al., {\it Biodemographic trajectories of
longevity}, Science 280, 855-860 (1998).
                                                                                
\bibitem{finch90} C. E. Finch, {\it Longevity, Senescence, and the Genome}, 
(Univ. of Chicago Press, 1990).

\bibitem{rose94} M.R. Rose, {\it Evolutionary biology of ageing}, (Oxford Univ. Press, 1994). 

\bibitem{muellerrose} L.D. Mueller and M.R. Rose, {\it Evolutionary theory predicts late-life mortality plateaus}, Proc. Natl. Acad. Sci, {\bf 93}, 15249-15253 (1996). 

\bibitem{moss} S. Moss de Oliveira, P.M.C. de Oliveira, D. Stauffer:
{\it Evolution, Money, War and Computers}, Teubner, Stuttgart and Leipzig 1999.
                                                                                
\bibitem{gavrilov} L.A. Gavrilov, N.S. Gavrilova, {\it The reliability 
theory of aging and longevity}, J. Theor. Biology 213, 527-545 (2001).
                                                                                
\bibitem{aviv} A. Aviv, D. Levy and M. Mangel, {\it Growth, telomere 
dynamics, and successful and unsuccessful human aging}. Mech. Ageing 
Dev. 124, 829-837 (2003); M. Masa, S. Cebrat and D. Stauffer, {\it 
Does telomere elongation in cloned organisms lead to a longer lifespan 
if cancer is considered?}, preprint, q-bio.PE/0408026. 

\bibitem{joeng} K.S. Joeng, E.J. Song, K.J. Lee and J. Lee, {\it Long 
lifespan in worms with long telomeric DNA}, Nature genetics, May 02, 2004. 

\bibitem{goel71a} N.S. Goel, S.C. Maitra and E.W. Montroll, {\it 
Nonlinear models of interacting populations} (Academic Press, 1971). 

\bibitem{goel71b} N.S. Goel, S.C. Maitra and E.W. Montroll, 
{\it On the Volterra and other nonlinear models of interacting 
populations}, Rev. Mod. Phys. {\bf 43}, 231 (1971).

\bibitem{pielou77} E.C. Pielou, {\it Mathematical ecology}, 
(Wiley, 1977).

\bibitem{emlen84} J.M. Emlen, {\it Population biology}, (Macmillan, 1984).

\bibitem{may74} R.M. May {\it Stability and complexity in model ecosystems}, 
(Princeton univ. Press, 1973).

\bibitem{svirezhev83} Yu. M. Svirezhev and D.O. Logofet, {\it Stability 
of biological communities} (Mir publishers, 1983). 

\bibitem{logofet93} D.O. Logofet, {\it Matrices and graphs: stability 
problems in mathematical ecology} (CRC press, 1993).

\bibitem{hofbauer98} J. Hofbauer and K. Sigmund, {\it Evolutionary 
games and population dynamics} (Cambridge University Press, 1998).

\bibitem{mccann00} K.S. McCann, {\it The diversity-stability debate}, 
Nature, {\bf 405}, 228-233 (2000).

\bibitem{sinha2} S. Sinha and S. Sinha, {\it Evidence of universality 
for the May-Wigner stability theorem for random networks with local 
dynamics} preprint (2004).


\bibitem{czaran} T. Czaran, {\it Spatio-temporal models of population 
and community dynamics}, (Chapman and Hall, 1998). 

\bibitem{bascompte} J. Bascompte and R.V. Sole, (eds.) {\it Modelling 
Spatiotemporal dynamics in ecology} (Springer, 1998). 

\bibitem{tilman} D. Tilman and P. Kareiva (ed.) {\it Spatial Ecology} 
(Princeton University Press, 1997).  

\bibitem{singh} B.K. Singh, J.S. Rao, R. Ramaswamy and S. Sinha, 
{\it The role of heterogeneity on the spatiotemporal dynamics of 
host-parasite metapopulation}, Ecol. Modeling, {\bf 180}, 435-443 
(2004).

\bibitem{tainaka89} K. Tainaka, {\it Stationary pattern of vortices 
or strings in biological systems: lattice version of the Lotka-Volterra 
model}, Phys. Rev. Lett. {\bf 63}, 2688-2691 (1989). 

\bibitem{satulovsky94} J.E. Satulovsky and T. Tome, {\it Stochastic 
lattice gas model for a predator-prey system}, Phys. Rev. E {\bf 49}, 
5073-5079 (1994). 

\bibitem{boccara94} N. Boccara, O. Roblin and M. Roger, {\it Automata 
network predator-prey model with pursuit and evasion}, Phys. Rev. E 
{\bf 50}, 4531-4541 (1994). 

\bibitem{frachebourg96} L. Frachebourg, P.L. Krapivsky and E. Ben-Naim, 
{\it Spatial organization in cyclic Lotka-Volterra systems}, Phys. 
Rev. E {\bf 54}, 6186-6200 (1996). 

\bibitem{lipowski99} A. Lipowski, {\it Oscillatory behavior in a lattice 
prey-predator system}, Phys. Rev. E {\bf 60}, 5179-5184 (1999).

\bibitem{antal01a} T. Antal, M. Droz, A. Lipowski and G. Odor, 
{\it On the critical behavior of a lattice prey-predator model}, 
Phys. Rev. E {\bf 64}, 036118 (2001). 

\bibitem{antal01b} T. Antal and M. Droz, {\it Phase transitions and 
oscillations in a lattice prey-predator model}, Phys. Rev. E {\bf 63}, 
056119 (2001). 

\bibitem{droz02} M. Droz and A. Pekalski, {\it Dynamics of polulations 
in a changing environment}, Phys. Rev. E {\bf 65}, 051911 (2002). 

\bibitem{droz04} M. Droz and A. Pekalski, {\it Population dynamics with 
or without evolution: a physicist approach},
Physica A, {\bf 336}, 84-92 (2004).

\bibitem{johst99} K. Johst, M. Doebli and R. Brandl, {\it Evolution 
of complex dynamics in spatially structured populations}, Proc. Roy. 
Soc. Lond. B {\bf 266}, 1147-1154 (1999).

\bibitem{pimm} S. L. Pimm, {\it Food Webs} (Chapman and Hall, London, 1982). 

\bibitem{polis} G.A. Polis and K. O. Winemiller, (eds.), {\it Food Webs: 
Integration of Patterns and Dynamics} (Chapman and Hall, New York, 1996). 

\bibitem{drossel03} B. Drossel, and A.J. McKane,  {\it Modelling food 
webs}, p.  218 in: {\it Handbook of Graphs and Networks - From the 
Genome to the Internet}, Bornholdt, S. and Schuster, H.G. (eds.), 
Wiley-VCH, Weinheim, 2003. 

\bibitem{cohen90} J.E. Cohen, T. Luczak, C. M. Newman and Z. -M. Zhou,  1990 
{\it Stochastic structure and nonlinear dynamics of food webs: 
qualitative stability in a Lotka-Volterra cascade model}, Proc. Roy. Soc. 
Lond. B {\bf 240}, 607-627 (1990). 

\bibitem{hall91} S.J. Hall and D. Raffaelli, {\it Food web patterns: 
lessons from a species-rich web}, J. Anim. Ecol. {\bf 60}, 823-842 (1991).

\bibitem{goldwasser93} L. Goldwasser and J. Roughgarden, {\it 
Construction and analysis of a large Carribean food web}, Ecology 
{\bf 74}, 1216-1233 (1993).

\bibitem{martinez95a} N.D. Martinez and J.H. Lawton, {\it Scale and 
food web structure- from local to global}, Oikos, {\bf 73}, 148-154 (1995).

\bibitem{martinez00} R.J. Williams and N.D. Martinez, {\it Simple 
rules yield complex food webs}, Nature {\bf 404}, 180-183 (2000). 

\bibitem{martinez02} J.A. Dunne, R.J. Williams and N.D. Martinez, 
{\it Food-web structure and network theory: the role of connectance 
and size}, Proc. Natl. Acad. Sci. {\bf 99}, 12917-12922 (2002). 

\bibitem{martinez02a} J.A. Dunne, R.J. Williams and N.D. Martinez, 
{\it Network structure and biodiversity loss in food webs: 
robustness increases with connectance}, Ecology Letters {\bf 5}, 
558-567 (2002).

\bibitem{martinez04} U. Brose, A. Ostling, K. Harrison and N.D. 
Martinez, {\it Unified spatial scaling of species and their trophic 
interactions}, Nature {\bf 428}, 167-171 (2004).

\bibitem{briand84} F. Briand and J.E. Cohen, {\it Community food 
webs have scale-invariant structure}, Nature, {\bf 307}, 264-266 
(1984).

\bibitem{cohenbook} J.E. Cohen, F. Briand and C.M. Newman, {\it 
Community food webs- data and theory}, Biomathematics, vol.20 
(Springer, 1990).

\bibitem{montoya01} R.V. Sole and J.M. Montoya, {\it Complexity and 
fragility in ecological networks}, Proc. Roy. Soc. Lond. B {\bf 268}, 
2039-2045 (2001).

\bibitem{montoya02} J.M. Montoya and R.V. Sole, {\it Small world 
patterns in food webs}, J. Theor. Biol. {\bf 214}, 405-412 (2002).

\bibitem{jennings03} S. Jennings and K.J. Warr, {\it Smaller 
predator-prey body size ratios in longer food chains}, Proc. Roy. 
Soc. Lond. B {\bf 270}, 1413-1417 (2003).

\bibitem{strogatz01} S.H. Strogatz, {\it Exploring complex networks}, 
Nature {\bf 410}, 268-275 (2001).

\bibitem{barabasi02} R. Albert and A.L. Barabasi, {\it Statistical mechanics of complex networks}, Rev. Mod. Phys. {\bf 74}, 47-97 (2002). 

\bibitem{schluter00} D. Schluter, {\it Introduction to the symposium: species 
interactions and adaptive radiation}, Am. Nat. {\bf 156}, S1-S3 (2000). 

\bibitem{diekmann99} U. Diekmann and M. Doebeli, {\it On the origin of 
species by sympatric spaciation}, Nature, {\bf 400}, 354-357 (1999).

\bibitem{doebeli00} M. Doebeli and U. Diekmann, {\it Evolutionary 
branching and sympatric speciation caused by different types of 
ecological interactions}, Am. Nat. {\bf 156}, S77-S101 (2000).

\bibitem{kirkpatrick02} M. Kirkpatrick and V. Ravigne, {\it Speciation 
by natural and sexual selection: models and experiments}, Am. Nat. 
{\bf 159}, S22-S35 (2002). 

\bibitem{gavrilets98} S. Gavrilets, H. Li and M.D. Vose, {\it Rapid 
parapatric speciation on holey adaptive lavdscapes}, Proc. Roy. Soc. 
Lond. B {\bf 265}, 1483-1489 (1998). 

\bibitem{gavrilets99} S. Gavrilets, {\it A dynamical theory of 
speciation on holey adaptive landscapes}, Am. Nat. {\bf 154}, 1-22 (1999). 

\bibitem{gavrilets00} S. Gavrilets, {\it Waiting time to paramatric 
speciation}, Proc. Roy. Soc. Lond. B {\bf 267}, 2483-2492 (2000).

\bibitem{gavrilets03} S. Gavrilets, {\it Models of speciation: 
what have we learned in 40 years?}, Evolution, {\bf 57}, 2197-2215 
(2003).

\bibitem{kaneko00} K. Kaneko and T. Yomo, {\it Sympatric speciation: 
compliance with phenotype diversification from a single genotype}, 
Proc. Roy. Soc. Lond. B {\bf 267}, 2367-2373 (2000).

\bibitem{baksneppen} P. Bak and K. Sneppen, {\it Punctuated equilibria 
and criticality in a simple model of evolution}, Phys. Rev. Lett. 
{\bf 71}, 4083-4086 (1993). 

\bibitem{paczuski96} M. Paczuski, S. Maslov and P. Bak, {\it Avalanche 
dynamics in evolution, growth and depinning models}, Phys. Rev. E 
{\bf 53}, 414-443 (1996).

\bibitem{newsnep96a} M.E.J. Newman and K. Sneppen, {\it Avalanches, 
scaling and coherent noise}, Phys. Rev. E {\bf 54}, 6226-6231 (1996).

\bibitem{newsnep96b} K. Sneppen and M.E.J. Newman, {\it Coherent noise, 
scale invariance and intermittency in large systems}, Physica D 
{\bf 110}, 209-222 (1997). 

\bibitem{kramer96} M. Kramer, N. Vande Walle and M. Ausloos, {\it 
Speciations and extinctions in a self-organizing critical model of 
tree-like evolution}, J. Phys. I (France), {\bf 6}, 599-606 (1996)

\bibitem{vande96} N. Vandewalle and M. Ausloos, {\it The screening of 
species in a Darwinistic tree-like model of evolution}, Physica D 
{\bf 90}, 262 (1996). 

\bibitem{head97} D.A. Head and G.J. Rodgers, Phys. Rev. E {\bf 55}, 
3312 (1997).

\bibitem{solerev} R.V. Sole, {\it Statistical mechanics of network 
models of macroevolution and extinction}, in: {\it Statistical 
Mechanics of Biocomplexity}, Lec. Notes in Phys. 217-250 (Springer, 1999).

\bibitem{soleman1} R. V. Sole and S.C. Manrubia, {\it Extinction and 
self-orgnized criticality in a model of large-scale evolution}, 
Phys. Rev. E {\bf 54}, R42-R45 (1996).

\bibitem{soleman2} R. V. Sole and S.C. Manrubia, {\it Criticality and 
unpredictability in macroevolution}, Phys. Rev. E {\bf 55}, 4500-4507 
(1997). 

\bibitem{manpac96} S.C. Manrubia and M. Paczuski, {\it A simple model of 
large scale organization in evolution}, Int. J. Mod. Phys. C {bf 9}, 
1025-1032 (1998).

\bibitem{solebas96} R.V. Sole and J. Bascompte, {\it Are critical 
phenomena relevant to large scale evolution?}, Proc. Roy. Soc. Lond. 
B {\bf 263}, 161-168 (1996). 

\bibitem{soleetal96} R.V. Sole, J. Bascompte and S.C. Manrubia, 
{\it Extinction: bad genes or weak chaos?}, Proc. Roy. Soc. Lond. B 
{\bf 263}, 1407-1413 (1996). 

\bibitem{soleetal97} R.V. Sole, S.C. Manrubia, M. Benton and P. Bak, 
{\it Self-similarity of extinction statistics in the fossil record}, 
Nature, {\bf 388}, 764-767 (1997); see also J.W. Kirchner and A. Weil, 
{\it No fractals in fossil extinction records}, Nature {\bf 395}, 
337-338 (1998).

\bibitem{roberts96} B.W. Roberts and M.E.J. Newman, {\it A model of 
evolution and extinction}, J. Theor. Biol. {\bf 180}, 39-54 (1996).

\bibitem{newman96} M.E.J. Newman, {\it Self-organized criticality, 
evolution and the fossil extinction record}, Proc. Roy. Soc. Lond. B 
{\bf 263}, 1605-1610 (1996).

\bibitem{newman97} M.E.J. Newman, {\it Evidence for self-organized 
criticality in evolution}, Physica D {\bf 107}, 293-296 (1997).

\bibitem{newman00} M.E.J. Newman, {\it Simple models of evolution and 
extinction}, Comp. in Sc. and Engg. {\bf 2}, 80-86 (2000).

\bibitem{wilke97} C. Wilke and T. Martinetz, 
{\it Simple model of evolution with variable system size}
Phys. Rev. E {\bf 56}, 7128-7131 (1997).


\bibitem{soc} H. J. Jensen, {\it Self-Organized Criticality : emergent 
complex behavior in physical and biological systems} (Cambridge University 
Press, 1998).

\bibitem{gould1} S.J. Gould and N. Eldredge, {\it Punctuated equilibria: 
the tempo and mode of evolution reconsidered}, Paleobiology {\bf 3}, 
115-151 (1977). 

\bibitem{gould2} S.J. Gould and N. Eldredge, {\it Punctuated equilibrium comes of age}, Nature {\bf 366}, 223-227 (1993).

\bibitem{boettcher} P. Bak and S. Boettcher, {\it Self-organized 
criticality and punctuated equilibria} Physica D {\bf 107}, 143-150 (1997).

\bibitem{kauffman} S. Kauffman, {\it The Origins of Order: 
Self-organization and selection in Evolution} (Oxford University Press, 
1993). 

\bibitem{peliti1} L  Peliti, {\it Fitness landscapes and evolution}, 
in: {\it Physics of Biomaterials: Fluctuations, Selfassembly and Evolution 
(NATO ASI series}, eds. T. Riste and D. Sherrington (Kluwer Academic 
Publishers, 1996) (also available at cond-mat/9505003 (1995)). 

\bibitem{peliti2} L. Peliti, {\it Introduction to the statistical theory 
of Darwinian evolution}, Lectures delivered at the ICTP Summer College 
on Frustrated Systems, Trieste, August, 1997 (available at 
cond-mat/9712027 (1997)). 

\bibitem{wilkerev} C.O. Wilke, C. Ronnewinkel, T. Martinez, {\it 
Dynamic fitness landscapes in molecular evolution} Phys. Rep. 
{\bf 349}, 395-446 (2001).

\bibitem{brookfield} J.F.Y. Brookfield, {\it Predicting the future}, 
Nature {\bf 411}, 999 (2001).

\bibitem{amaral} L.A.N. Amaral and M. Meyer, {\it Environmental changes, 
coextinction, and patterns in the fossil record}, Phys. Rev. Lett. 
{\bf 82}, 652-655 (1999). 

\bibitem{camacho} J. Camacho and R.V. Sole, {\it Extinction and taxonomy 
in a trophic model of coevolution}, Phys. Rev. E {\bf 62}, 1119-1123 (2000).

\bibitem{drossel98} B. Drossel, {\it Extinction events and species 
lifetimes in a simple ecological model}, Phys. Rev. Lett. {\bf 81}, 
5011-5014 (1998).

\bibitem{sinha} C. Wilmers, S. Sinha and M. Brede, {\it Examining 
the effects of species richness on community stability: an assembly 
model approach}, Oikos {\bf 99}, 363-367 (2002). 

\bibitem{lloyd93} E.A. Lloyd and S.J. Gould, {\it Species selection 
on variability}, Proc. Natl. Acad. Sci.  {\bf 90}, 595-599 (1993). 

\bibitem{mayr97} E. Mayr, {\it The objects of selection}, Proc. Natl. 
Acad. sci. {\bf 94}, 2091-2094 (1997)

\bibitem{gould99} S.J. Gould and E.A. Lloyd, {\it Individuality and adaptation across levels of selection: how shall we name and generalize the unit of Darwinism?}, Proc. Natl. Acad. Sci. {\bf 96}, 11904-11909 (1999).

\bibitem{johnson02} C.R. Johnson and M.C. Boerlijst, {\it Selection at 
the level of the community: the importance of spatial structure}, 
Trends in ecology and evolution, {\bf 17}, 83-90 (2002). 

\bibitem{thompson98a} J.N. Thompson, {\it Rapid evolution as an 
ecological process}, Trends in Ecol. Evol. {\bf 13}, 329-332 (1998). 

\bibitem{thompson98b} J.N. Thompson, {\it The population biology 
of coevolution}, Res. Popul. Ecol. {\bf 40}, 159-166 (1998).

\bibitem{thompson99} J.N. Thompson, {\it The evolution of species 
interactions}, Science {\bf 284}, 2116-2118 (1999).

\bibitem{stockwell03} C.A. Stockwell, A.P. Hendry and M.T. Kinnison, 
{\it Contemporary evolution meets conservation biology}, Trends in 
Ecology and Evolution, {\bf 18}, 94-101 (2003).

\bibitem{turchin03} P. Turchin, {\it Ecology: evolution in population 
dynamics}, Nature, {\bf 424}, 257-258 (2003). 

\bibitem{yoshida03} T. Yoshida, L.E. Jones, S.P. Ellner, G.F. 
Fussmann and N.G. Hairston Jr., {\it Rapid evolution drives 
ecological dynamics in a predator-prey system}, Nature, {\bf 424}, 
303-306 (2003).

\bibitem{fussmann03} G. F. Fussmann, S.P. Ellner, and N.G. Hairston, Jr. 
{\it Evolution as a critical component of plankton dynamics}, Proc. 
Roy. Soc. Lond. B {\bf 270}, 1015-1022 (2003). 

\bibitem{martinez95} N. D. Martinez, {\it Unifying ecological 
subdisciplines with ecosystem food webs}, in: {\it Linking 
species and ecosystems}, eds. C.G. Jones and J.H. Lawton 
(Chapman and Hall, 1995).

\bibitem{abramson} G. Abramson, {\it Ecological model of extinctions}, 
Phys. Rev. E {\bf 55}, 785-788 (1997).


\bibitem{mckane98} G. Caldarelli, P.G. Higgs and A.J. McKane, 
{\it Modelling coevolution in multispecies communities},
J. Theor. Biol. {\bf 193}, 345-358 (1998); preprint (2004). 

\bibitem{mckane01} B. Drossel, P.G. Higgs and A.J. McKane, 
{\it The influence of predator-prey population dynamics on the 
long-term evolution of food web structure},
J. Theor. BIol. {\bf 208}, 91 (2001). 


\bibitem{quince1} C. Quince, P.G. Higgs and A.J. McKane, {\it Deleting 
species from model food webs}, q-bio.PE/0402014 at e-print server 
www.arXiv.com.. 

\bibitem{quince2} C. Quince, P.G. Higgs and A.J. McKane,{\it Topological 
structure and interaction strengths in model food webs}, q-bio.PE/0401037, 
at e-print server www.arXiv.com.


\bibitem{chowstauprl} D. Chowdhury, D. Stauffer and A. Kunwar, {\it Unification of small and large time scales for biological evolution: deviations from power law},  Phys. Rev. Lett. 90, 068101 (2003).                                                                                
\bibitem{chowstaupre} D. Chowdhury and D. Stauffer, {\it Food-web-based unified model of macro- and microevolution},   Phys. Rev. E 68, 041901 (2003).
                                                                                
\bibitem{stauchowpa} D. Stauffer and D. Chowdhury, {\it Unified "micro"- and "macro-" evolution of eco-systems: Self-organization of a dynamic network},  Physica A, 336, 102-111 (2004).
                                                                                
\bibitem{chowstaupa} D. Chowdhury and D. Stauffer, {\it Computer simulations of history of life: speciation, emergence of complex species
from simpler organisms, and extinctions}, Physica A, {\bf 340}, 685-696 (2004).                                                                                
\bibitem{csnetofnet} D. Chowdhury and D. Stauffer, {\it Evolving eco-system: a network of networks}, Physica A {\bf 346}, 68-74 (2005).

\bibitem{stauetal04} D. Stauffer, A. Kunwar and D. Chowdhury, {\it 
Evolutionary ecology in-silico:evolving foodwebs, migrating population 
and speciation}, Physics A (in press, 2005).                                                               
\bibitem{rikvold03} P.A. Rikvold and R.K.P. Zia, {\it Punctuated equilibria and 1/f noise in a biological coevolution model with individual-based dynamics},  Phys. Rev. E {\bf 68}, 031913 (2003). 

\bibitem{hall1} M. Hall, K. Christiensen, S.A. di Collobiano and H.J. Jensen, {\it Time-dependent extinction rate and species abundance in a tangled-nature model of biological evolution}, Phys. Rev. E {\bf 66}, 011904 (2002). 

\bibitem{collobiano1} S. A. di Collobiano, K. Christiensen and H.J. Jensen, {\it The tangled nature model as an evolving quasi-species model}, J. Phys. A {\bf 36}, 883-891 (2003). 

\bibitem{rama} S. Singh and R. Ramaswamy, {\it Comment on ``unification 
of small and large time scales for biological evolution: deviations 
from power law''}, private communication (2004).
                                                                                

\end{thebibliography}
\end{document}